**Zirconium oxidation under high energy heavy ion irradiation**

N. Bérerd, A. Chevarier, N. Moncoffre, H. Jaffrezic,
Institut de Physique Nucléaire de Lyon ,4, rue Enrico Fermi , F-69622 Villeurbanne Cedex, France,
E. Balanzat,
CIRIL, CEA-CNRS-ENSICAEN, rue Claude Bloch, BP 5133, F-14070 Caen Cedex 5, France.
H. Catalette,
EDF R&D, Site des renardières, 77818 Moret sur Loingt, Cedex, France.

Abstract

This paper concerns the study of zirconium oxidation under irradiation with high energetic Xe ions. The irradiations were performed on the IRRSUD beam line at GANIL (Caen). The oxygen partial pressure was fixed at $10^{-3}$ Pa and two temperature conditions were used, either 480°C reached by Joule effect heating or 280°C due to Xe energy deposition. Zirconia was fully characterized by Rutherford Backscattering Spectrometry, Transmission Electron Microscopy and Grazing Angle X-ray Diffraction. Apparent diffusion coefficients of oxygen in $ZrO_2$ were determined from these experiments by using a model which takes into account a surface exchange between oxygen gas and the $ZrO_2$ surface. These results are compared with thermal oxidation data.

# I. INTRODUCTION

This paper deals with the influence of heavy ion irradiation on the oxidation kinetics of zirconium. This study is performed in the framework of long-term storage of nuclear spent fuel. In France, Zircaloy is used as cladding material in Pressured Water Reactors. During reactor operation, it is well established that the oxidation process occurs on both sides of the zircaloy tube. Whereas the outer side of the tube is oxidized in contact with water, the inner surface is oxidised in a dry atmosphere. Evermore the fuel cladding inner surface is submitted to neutrons, alpha, gamma and fission product irradiation. The characteristics of the so formed oxide will influence the long term behaviour of the spent fuel assembly in case of direct storage without reprocessing.

The corrosion behaviour of zirconium alloys as cladding materials has been so far discussed for many years. A number of review papers has been published on this subject[1,2]. It has been established that thermal oxidation kinetics of zirconium and zircaloy at atmospheric pressure obeys a parabolic or cubic law preceding the linear one. In the pretransition step, the oxide film growth mechanism is the migration of oxygen through the corrosion film with the formation of a new oxide at the metal-oxide interface[3]. This mechanism has been modelled and compared to experimental data[4]. Let us remind that the zirconia crystallographic structure presents three polymorphs of $ZrO_2$. At atmospheric pressure zirconia is monoclinic below 1300 K. Around 1400 K, a phase transition occurs and the crystal takes a tetragonal structure. Near 2600 K a second phase transition leads to the cubic structure. In the 500-1100 K temperature range, the crystalline structure of the oxide, formed in the pretransition state during thermal oxidation, is monoclinic with a proportion of tetragonal phase stabilized by local stress fields[5,6]. Although irradiation effects with heavy ions in pure monoclinic or doped zirconia have drawn attention[7,8,9,10], up to our knowledge available data on radiation effects on zirconia formation are restricted to neutron[11] and electron[12] irradiations.

Using the reactor of the Institut Laue Langevin in Grenoble (ILL), we have recently observed an increase of the oxidation rate mainly induced by fission product irradiation[13]. The experiments of oxidation under irradiation performed with the high neutron flux reactor at ILL used the following conditions: a pressure in the H9 beam line of $5 \times 10^{-3}$ Pa, a temperature of 480°C, a neutron flux of $5 \times 10^{14}$ n cm$^{-3}$ s$^{-1}$ and the fission rate was about $10^{11}$ s$^{-1}$. The oxidation rate was measured by fission fragment spectroscopy on line. At the end of the experiment, after an irradiation time of 122 hours, it was shown that the whole Zr foil was completely oxidised corresponding to an oxide thickness of 3.1 µm and a stoichiometry close to $ZrO_2$. The determination of this stoichiometry is given with an accuracy of approximately 10%.

As post-irradiation characterisation of the so formed oxide was not possible due to a very high radioactivity, the results needed to be confirmed by further experiments. For that purpose, we have chosen to study the zirconium oxidation kinetics under heavy ion irradiation. This choice was supported by Hj. Matzke results which have experimentally demonstrated that enhancement of self-diffusion in $UO_2$ pellets in reactor conditions was mainly due to the defects and the morphologic changes induced by slowing down of fission products[14,15].

The aim of the present paper is to model the zirconium oxidation kinetics, under fission product irradiation only, represented by 0.5 MeV/u xenon ions. The irradiations were performed on GANIL accelerator in Caen. The oxidation rate under irradiation is compared to thermal data obtained in the same pressure and temperature conditions in order to evaluate the damage role. The evolution of the crystallographic structure is also discussed.

## II. RADIATION ENHANCED OXIDATION UNDER Xe IRRADIATION

### A. Experimental procedure

#### i) Irradiation conditions

The irradiations were performed on the IRRSUD beam line at GANIL accelerator. A $^{129}$Xe beam was used at a kinetic energy of 64.5 MeV and at a flux of $2.6 \times 10^{10}$ Xe cm$^{-2}$ s$^{-1}$. An irradiation cell was specifically realized for the experiments, which allows to control precisely pressure and temperature (figure 1a). As shown in figure 1b, the target is made of 6 µm thick zirconium foils which are positioned on a copper support. A tantalum resistance can heat the target by Joule effect up to 500°C. To maintain the resistance electric isolation, a mica plate is inserted between copper and tantalum. The whole set is fixed on stainless steel on which is placed thermocouple. A variable power supply allows to control the intensity into the resistance, and to fix the target temperature. Two experimental conditions were used either by heating the zirconium foil at 480°C or without exterior thermal heating. In this last condition, the ion energy deposition leads to a target temperature of 280°C.

The pressure of $5 \times 10^{-3}$ Pa was controlled with a microleak and a 1.5 µm thick aluminum window was placed between the irradiation cell and the IRRSUD beam line in order to avoid any pressure degradation in the beam line. The Xe beam kinetic energy after crossing the Al window is 50 MeV which corresponds to a Xe range in Zr equal to 5.7 µm.

The zirconium foils were irradiated during different times up to 47 hours which allows to vary the studied fluences between 0.25 and $4 \times 10^{15}$ Xe cm$^{-2}$.

#### ii) X-ray Diffraction analysis

Grazing angle X-ray diffractometry was used to obtain the volumic proportion of monoclinic and tetragonal zirconia phases induced as a function of the xenon fluence. Measurements were performed at the Ecole Centrale of Lyon. The incident X-ray beam angle was 2° which allows the analysis of the first 600 nm. The volume fraction C of the tetragonal zirconia phase is obtained by measuring the intensity of the tetragonal (111) ray (31.5°) and of the monoclinic ($\bar{1}11$) ray (28.17°), according to [16,17]:

$$C = \frac{I_t(111)}{I_t(111) + 2.042 I_m(\bar{1}11)} \quad (1)$$

where $I_t(111)$ is the (111) ray intensity and $I_m(\bar{1}11)$ is the ($\bar{1}11$) ray intensity.

**B. Experimental results**

For each temperature and fluence, the oxide layer thickness was determined using Rutherford Backscattering Spectrometry using α particles between 2 and 6 MeV energies on the 4 MV Van de Graaff accelerator of the IPNL.

Figure 2 displays two typical RBS spectra obtained on a Zr sample irradiated at 480°C during 5 and 13 hours. By fitting the experimental RBS spectra, we deduced first the stoichiometry of the formed oxide and the oxide thickness. We found a $ZrO_2$ stoichiometry and 270 and 680 nm thickness respectively for 5 and 13 hours irradiation times. Secondly, this technique allows to determine the oxygen gain G for each irradiation time and consequently, the oxidation kinetics of zirconium for both temperatures. Tables 1 and 2 summarize all the results obtained as a function of fluence on samples irradiated at 480 and 280°C.

The image obtained by Scanning Electron Microscopy (SEM) of the sample irradiated at the highest fluence of $4.4 \times 10^{15}$ Xe cm$^{-2}$ (figure 3), clearly shows that the oxidation occurs only on the face regarding the beam and that the sample is completely dense and uniform even after the longest irradiation. This last observation seems to prove that the oxide layer remains in a pre-transition phase.

In order to characterize the different zirconia structures formed as a function of the irradiation conditions, a grazing angle X-ray diffraction (GAXRD) analysis was performed.

Whereas most authors have investigated phase transformation of zirconia under irradiation[7-10], the originality of this work is to study the irradiation induced effects on the as-growing zirconia structural evolution. An illustration of obtained X-ray spectra is presented in figure 4. From the relation (1), the relative volumic percentages of tetragonal phase have been calculated for each fluence and temperature. They are reported in tables 1 and 2 and plotted in figure 5. At 280°C, only the tetragonal phase is observed whereas, at 480°C, a mixture of tetragonal and monoclinic phases is found (50% of each already for the $1.2 \times 10^{15}$ Xe cm$^{-2}$ fluence).

We have calculated, from the plot of Ln G as a function of Ln t, that the time dependence of the oxidation kinetics law is in $t^{0.88}$ and $t^{0.75}$ respectively at 480 and 280°C. It shows that we are not in the classical case of $\sqrt{t}$ dependence, observed in case of thermal oxidation. An illustration is given in figure 6 which represents the oxidation kinetic law at 480°C. As the oxidation kinetics is linear above an irradiation time of 10 hours, we deduced in first approximation oxidation rates equal to 1.9 and $2.1 \times 10^{17}$ O at. cm$^{-2}$ h$^{-1}$ respectively at 280 and 480°C.

From these observations, a model will be developed further in this paper, which allows to fit the oxygen gain under irradiation.

**III. THERMAL OXIDATION STUDY**

In previous works[13, 18, 19], we have studied thermal oxidation at reduced ($P_{O2} = 10^{-3}$ Pa) and atmospheric ($P_{O2} = 2 \times 10^4$ Pa) pressures. The annealing conditions (time and temperature range) were chosen in order to study the pre-transition oxidation kinetics for which the mechanism of oxide growth is known to be related to oxygen migration through the oxidation film. The sample weight gain measurements were performed by the thermogravimetry method and RBS. The details of the experimental procedure are given in references[18, 19].

The apparent oxygen diffusion coefficients D were deduced from the measurements, assuming that the oxide growth is limited by a pure oxygen diffusion process through the existing oxide layer. The temperature dependence of the diffusion coefficient is thus:

$$D = D_0 \exp(-\frac{E_a}{k.T}) \quad (2)$$

where $D_0$ is the pre-exponential factor, k is the Boltzmann constant ($8.617 \times 10^{-5}$ eV.K$^{-1}$), T the temperature, $E_a$ the activation energy (eV.at$^{-1}$). The $D_0$ and $E_a$ values for each pressure are obtained by the least square fitting of the Arrhenius plot. The results are summarized in table 3. In addition, the sample structure was studied by X ray diffractometry. This technique showed that, in the 350-480°C temperature range, the growing zirconia layer consisted in monoclinic (80 %) and tetragonal (20%) zirconia.

Our results have been compared with data from the literature[20-27] (figure 7). For this comparison we have selected data obtained in tetragonal and monoclinic zirconia at atmospheric pressure. Diffusion coefficients in tetragonal zirconia have been measured by using the tracer technique while in monoclinic zirconia, most data have been obtained from oxidation growth experiments. Figure 7 puts forward the following points:

i) There is a continuity between oxygen thermal diffusivity in tetragonal (dashed lines) and monoclinic (full lines and dots) zirconia.

ii) The slopes of the Arrhenius plots are close one to each other, whatever the crystallographic structure. The deduced activation energies stand around 1.1 ± 0.3 eV at$^{-1}$. This value is interpreted as oxygen diffusion through oxygen vacancies in the $ZrO_2$ lattice.

iii) Concerning our data obtained at two oxygen partial pressures different of seven orders of magnitude, the measured diffusion coefficients vary only by one order of magnitude. From the two measurements at 450°C, we have deduced a dependency in $P_{O2}^{0,14}$ in agreement with the literature[28, 29].

## IV. RADIATION ENHANCED OXIDATION STUDY

Experimental results under irradiation differ from thermal results by two main points:

(i) the non-$\sqrt{t}$ diffusion dependence,

(ii) a surface oxygen concentration which is time dependant due to the irradiation process.

Hence, we propose to consider both the oxygen radiation enhanced diffusion $D^*$ and the reaction constant K (cm.s$^{-1}$) at the surface. In this approach, the Fick's second law resolution is[3, 30]:

$$\frac{C-Co}{C_S^0-Co} = \text{erfc}\left(\frac{x}{2\sqrt{D^*.t}}\right) - \exp(h.x+h^2.D^*.t).\text{erfc}\left(\frac{x+2.h.D^*.t}{2\sqrt{D^*.t}}\right) \quad (3)$$

with: C(at.cm$^{-3}$) : Oxygen concentration in the zirconium at a depth x and a time t.

Co (at.cm$^{-3}$) : Oxygen concentration in the zirconium in the bulk at t = 0.

$C_S^0$ (at.cm$^{-3}$) : Oxygen surface concentration at equilibrium. It is deduced from ZrO$_2$ stoichiometry, it is equal to 5.5x10$^{22}$ at.cm$^{-3}$.

h = $\frac{K}{D^*}$ (cm$^{-1}$).

We used this expression in order to calculate the evolution of oxygen gain, G(t), as a function of time:

$$G(t) = \int_0^t Jo.dt$$

where Jo is the oxygen flux through the surface.

From the Fick's law,

$$Jo = -D^*.\frac{\partial C}{\partial t} \quad (4)$$

The expression (3) has been derived as function of t, and the concentration Co in the bulk taken equal to 0. At the surface, x = 0, this leads to the following expression:

$$G(t) = \int_0^t Jo.dt = \frac{C_s^0}{h}\left[\exp(h^2.D^*.t).erfc(h\sqrt{D^*.t})-1+\frac{2}{\sqrt{\pi}}.h\sqrt{D^*.t}\right] \quad (5)$$

The correct D* and h values are obtained by fitting the result of the model to the experimental kinetics using the MINUIT programme[31]. An illustration of such a simulation is presented in figure 8. The values of D* and K are given in table 4.

K is a kinetics constant characteristic of a surface exchange between residual oxygen gas and the zirconia solid surface. We found K values of the order of $10^{-9}$ cm s$^{-1}$ for 280 and 480°C. This value is comparable to that given by Kilo et al.[32] and Tannhauser et al.[33] for thermal isotopic oxygen exchange at the zirconia surface in the 500-700°C temperature range. This last author even underlined the very low dependence of K with temperature.

## V. DISCUSSION

The Arrhenius plots corresponding to oxygen diffusion both thermal and under irradiation are presented in figure 9 in the 280-480°C temperature range. As $D^* = D_{irr} + D_{th}$[3], D* is very close to $D_{irr}$ (see table 4). In addition, taking into account the errors, around 50%, deduced from MINUIT programme, the D* values are similar at 480 and 280°C. It puts forward that the enhanced oxidation due to defects created by the Xe ions is not thermally activated in this temperature range.

Despite the predominance of the electronic stopping power, we make the hypothesis that the diffusion process under irradiation is enhanced by the displacement cascades. We have estimated the oxygen diffusion coefficient in $ZrO_2$, $D_{irr}$, generated by Xe ions as usually considered by the Einstein relation[34]:

$$D_{irr} = \frac{1}{6} R^2 \Gamma \qquad (6)$$

where R is the root-mean square displacement of an oxygen atom in the collision cascade, $\Gamma$ is the jump rate proportional to the atomic displacement rate F in dpa s$^{-1}$ ($\Gamma = \alpha F$, where $\alpha$ corresponds to the number of atomic jumps per displacement).

From SRIM calculation[35], we deduced: R = 390 nm and F = 3x10$^{-5}$ dpa s$^{-1}$.

The $D_{irr}$ experimental values were reproduced by using equation (6) in which $\alpha$ is equal to 135 or 87 respectively at 480°C and 280°C. These rather high values are comparable to Müller's ones[34] who conclude to a significant influence of the collision cascade.

## VI. CONCLUSION

We have measured the zirconium oxidation rate under 50 MeV Xe irradiation in order to simulate the fission product energy slowing down in nuclear fuel claddings. These data have been interpreted by using a model developed by Crank[30]. Thus apparent oxygen diffusion coefficients enhanced by irradiation have been deduced as well as oxygen surface exchange constants. In the studied temperature range, the comparison between D* and thermal D values put forwards two important results: first, the oxygen diffusivity in $ZrO_2$ under Xe irradiation is athermal, second, it is strongly enhanced by a factor 6x10$^4$ at 280°C.

**Acknowledgements**

The authors would like to thank very sincerely N. Chevarier and J.C. Duclot for their precious help during this work. They are also grateful to L. Maunoury and J. M. Ramillon for their technical support during GANIL experiments.

References

[1] C.Lemaignan, A.T. Motta; Zirconium alloys in Nuclear Applications in Materials Science and technology, R.W. Cahn, P. Haasen, E.J. Kramer (Ed) 1994 vol 10B.

[2] B. Cox, Oxidation of zirconium and its alloys in Advance in Corrosion Science and Technology, Fontana Staeble (Ed) (1976), 173, vol.5.

[3] J. Philibert, Diffusion et transport de matière dans les solides, Les éditions de la Physique (Ed) 1985.

[4] E.A.Garcia ; J. Nucl. Mater 224 (1995) 299-304

[5] P. Barberis; J. Nucl. Mater 226 (1995) 34-43

[6] M. Parise, R. Foerch, G. Cailletaud ; Journal de Physique 9 (1999) 311-315

[7] K.E. Sickafus, H.J.Matzke, Th. Hatmann, K. Yasuda, J.A. Valdez, P. Chodak III, M. Nastasi, R.A. Verrall, J. Nucl. Mater 274 (1999) 66-77

[8] D. Simeone, J.L. Bechade, D. Gosset, A.Chevarier, P. Daniel, H. Pillaire, G. Baldinozzi, J. Nucl. Mater 281 (2000) 171-181

[9] D. Simeone, D. Gosset, J.L. Bechade, A.Chevarier; J. Nucl. Mater 300 (2002) 27-38

[10] A. Benyagoub, F. Couvreur, S. Bouffard, F. Levesque, C. Dufour, E. Paumier, Nucl. Instrum. and Meth.in Phys.Res.B 175-177 (2001) 417-421

[11] F. Lefebvre, C. Lemaignan ; J. Nucl. Mater 248 (1997) 268-274

[12] M.M.R. Howlader, C. kinoshita, K. Shiiyama, M. Kutsuwada, M.Inagaki, J. Nucl. Mater 265 (1999) 100-107

[13] N. Bérerd, H. Catalette, A. Chevarier, N. Chevarier, H. Faust, N. Moncoffre, Surface and Coating Technology 158-159 (2002) 473.

[14] A. Höh and Matzke; J. of Nucl. Mat. 48 (1973) 157-160

[15] Hj. Matzke; Rad. Effects 75 (1983) 317-319

[16] B. D. Cullity, Elements of X-ray diffraction, 2d Edition, Addison-Wesley Publishing Co, 1978.

[17] M. Parise, PhD thesis, Ecole Nationale Supérieure des Mines de Paris, 1996.

[18] N. Bérerd, Phd thesis, Lyon University, 2003

[19] K. Poulard, PhD thesis, Lyon University, 2001

[20] H.-I. Yoo, B.-J. Koo, J.-O. Hong, I. –S. Hwang, Y. –H. Jeong; J. of Nucl. Mat. 299 (2001) 235-241

[21] B. Cox, J. P. Pemsler; J. of Nucl. Mat. 28 (1968) 73-78

[22] B. Cox; J. of Nucl. Mat. 218 (1995) 261-264

[23] A. Grandjean, J. Kovacs; J. of Nucl. Mat. 210 (1994) 78-83

[24] B. Oberlander, P. Kofstad, I. Kves; Materialwissenschaft-und-werkstofftechnik 19(6) (1988) 190-193

[25] K. Park, D. R. Olander; J. Electrochem. Soc. 138(4) (1991) 1154-1159

[26] E. Y. Zamalin, A. G. Rakoch, T. V. Popova; Physics and chemistry of materials treatment 30(3) (1996) 248-252

[27] B.-K. Kim, S. J. Park, H. Hamaguchi; J. Am. Ceram. Soc. 77(10) (1994) 2648-2652

[28] F. J. Keneshea, D. L. Douglass, Oxidation metals, vol.3 (1971) 1-14.

[29] J. Nakamura, M. Hashimoto, T. Otomo, S. Kawasaki, J. of Nucl. Mat. 200 (1993) 256-264.

[30] J. Crank, The mathematics of diffusion, $2^{nd}$ edition, Clarendon Press (Ed), 1995.

[31] F. James, MINUIT, Function Minimization and Error Analysis, CERN Program Library Long Writeup D506

[32] M. Kilo, C. Argirusis, G. Borchardt, R. A. Jackson, Phys. Chem. Chem. Phys. 5 (2003) 2219.

[33] D. S. Tannhauser, J. A. Kilner, B. C. H. Steele, Nucl. Instr. And Meth. in Phys. Res. 218 (1983) 504.

[34] A. Müller, V. Nondorf, M.-P. Match, J. Appl. Phys. 64(7) (1998) 3445-3455.

[35] J. F. Ziegler, J. P. Biersack, U. Littmark; *The stopping and range of ions in solids*; Pergamon press, New York; 1985.

Figure captions

Figure 1: Irradiation set up allowing the pressure and temperature control: detail of the irradiation cell (a) and detail of the target holder (b).

Figure 2: RBS spectra obtained on 480°C annealed samples during 5 and 13 hours using 4 MeV α particles.

Figure 3: Cross-section SEM micrograph obtained with backscattered electrons on a sample irradiated with $4.4 \times 10^{15}$ Xe.cm$^{-2}$.

Figure 4: GAXRD spectra of samples irradiated during 10 hours without external heating (a) and samples irradiated during 13 hours at 480°C (b).

Figure 5: Evolution of the tetragonal phase in irradiated zirconia as a function of the Xe fluence.

Figure 6: Zirconium oxidation kinetics under irradiation at 480°C. The error bars are deduced from RBS measurements performed at increasing energies, depending on the oxide thickness.

Figure 7: Arrhenius plots of the oxygen thermal diffusion into zirconia: comparison between literature data and our work. (1-6 in monoclinic zirconia and 7-10 in tetragonal zirconia).

Figure 8: Zirconium oxidation kinetics under irradiation at 480°C. Comparison between experimental data (dots) and the model deduced from eq. 5 (full line).

Figure 9: Arrhenius plots of the oxygen diffusion in zirconia for thermal annealing and under irradiation.

Table captions

Table 1: Oxygen gain, oxide thickness and percentage of tetragonal phase as function of time or fluence for samples irradiated at 480°C.

Table 2: Oxygen gain, oxide thickness and percentage of tetragonal phase as function of time or fluence for samples irradiated without external heating.

Table 3: Oxygen thermal diffusion coefficients as a function of temperature. $D_0$ and $E_a$ values are also presented.

Table 4: D and K values deduced from the diffusion model (eq.5).

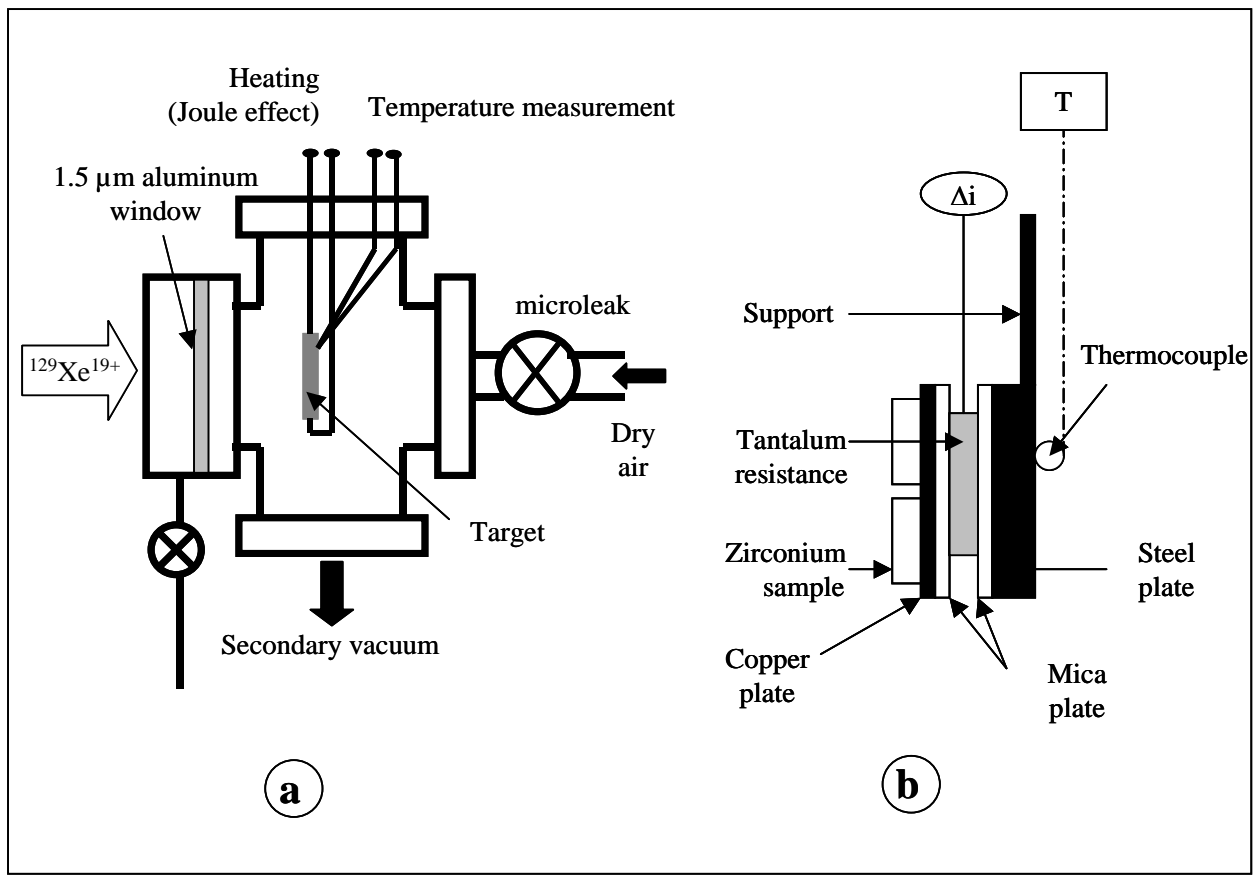

Figure 1

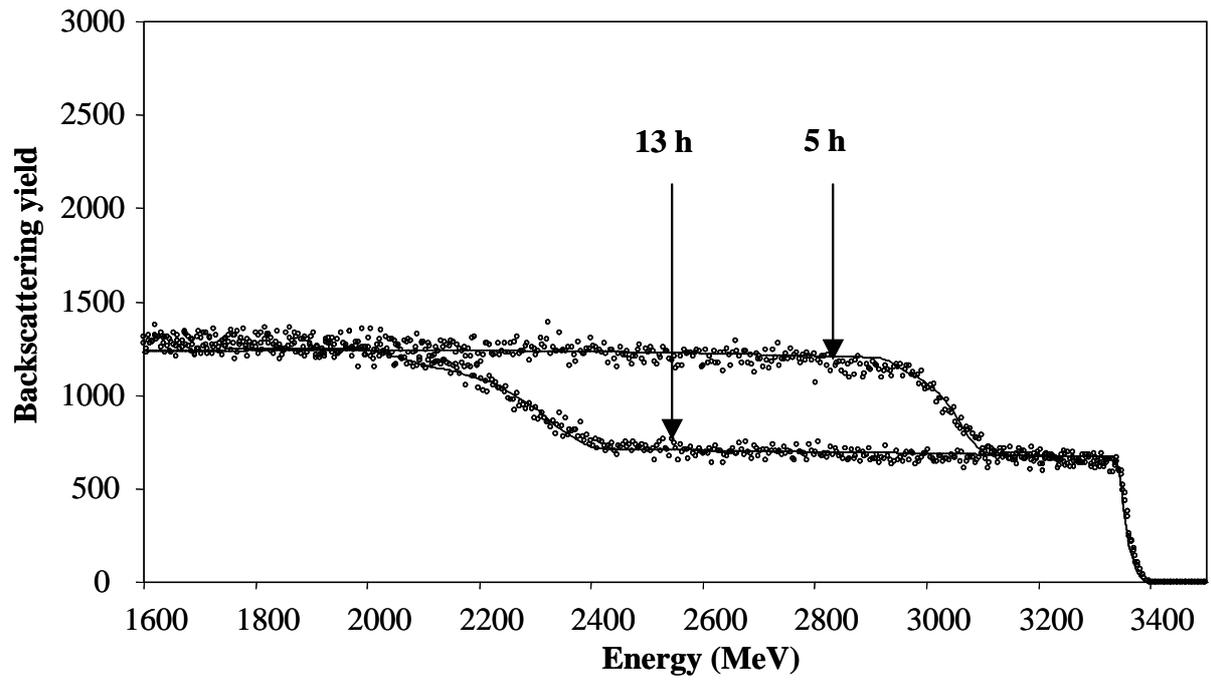

Figure 2

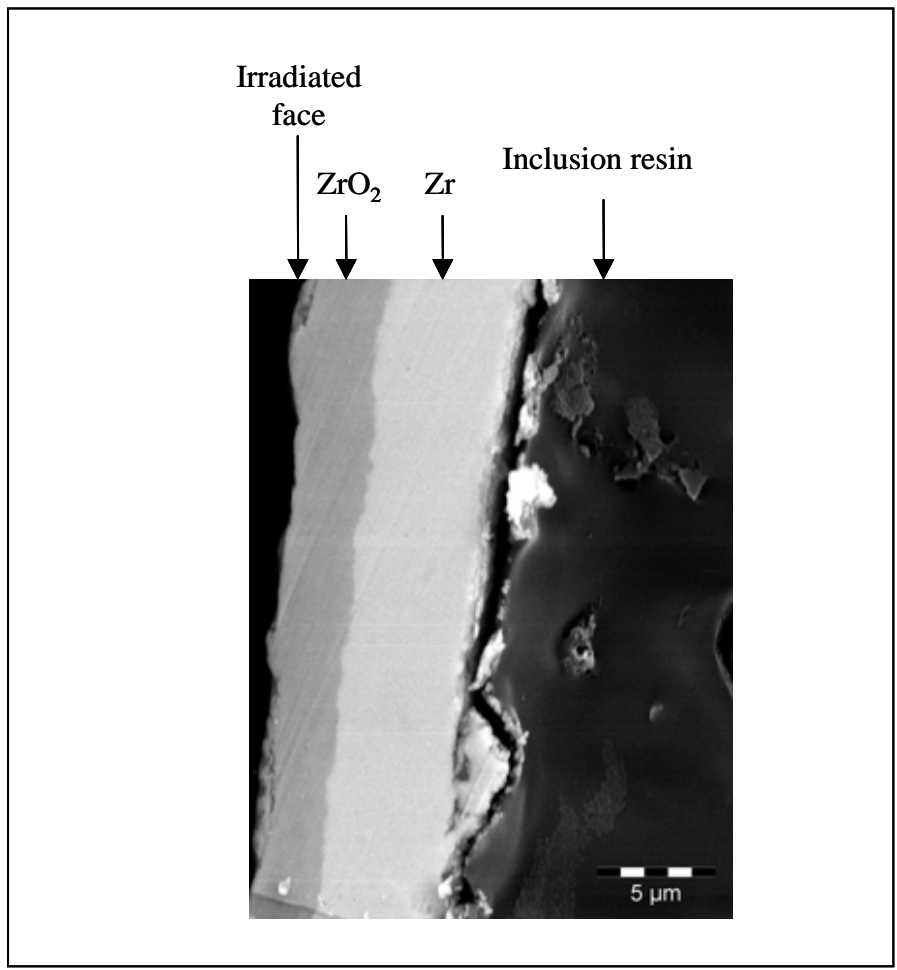

Figure 3

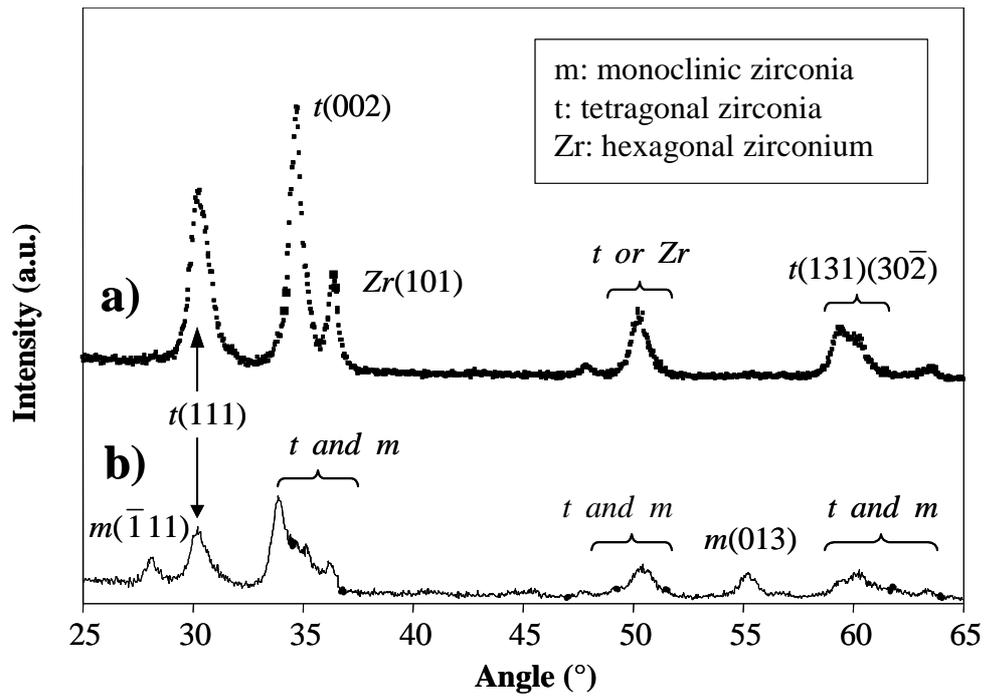

Figure 4

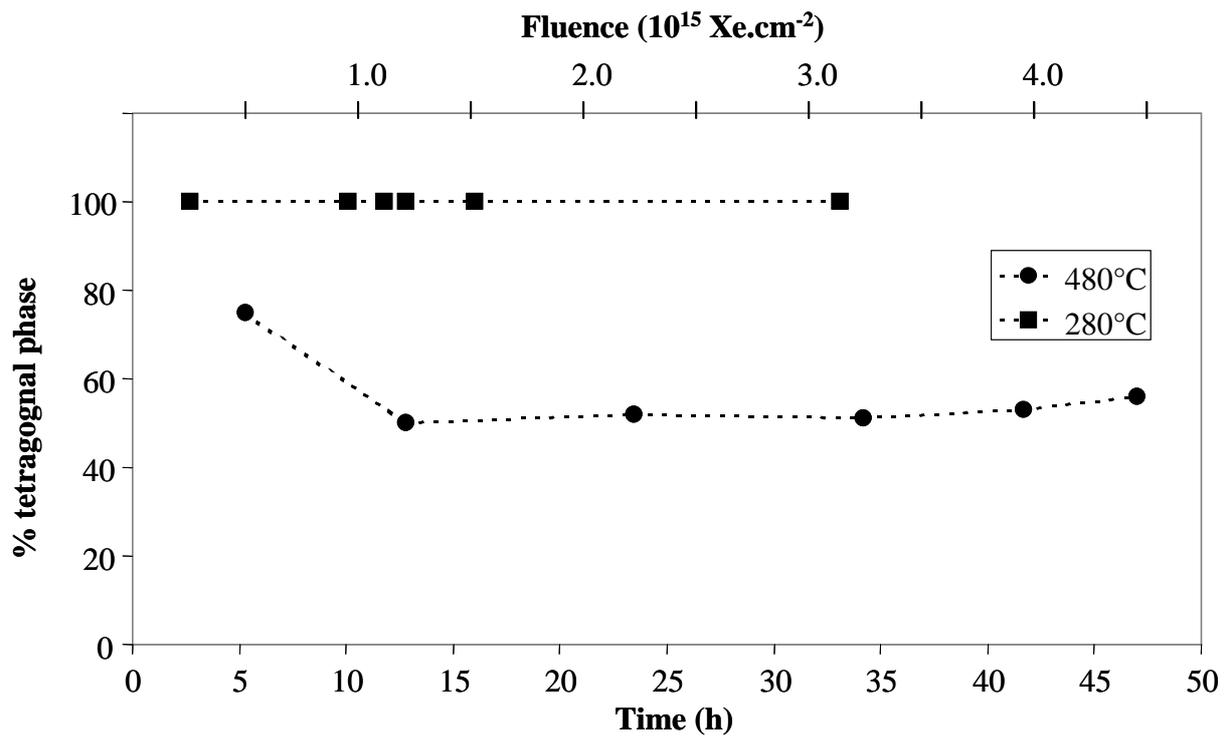

Figure 5

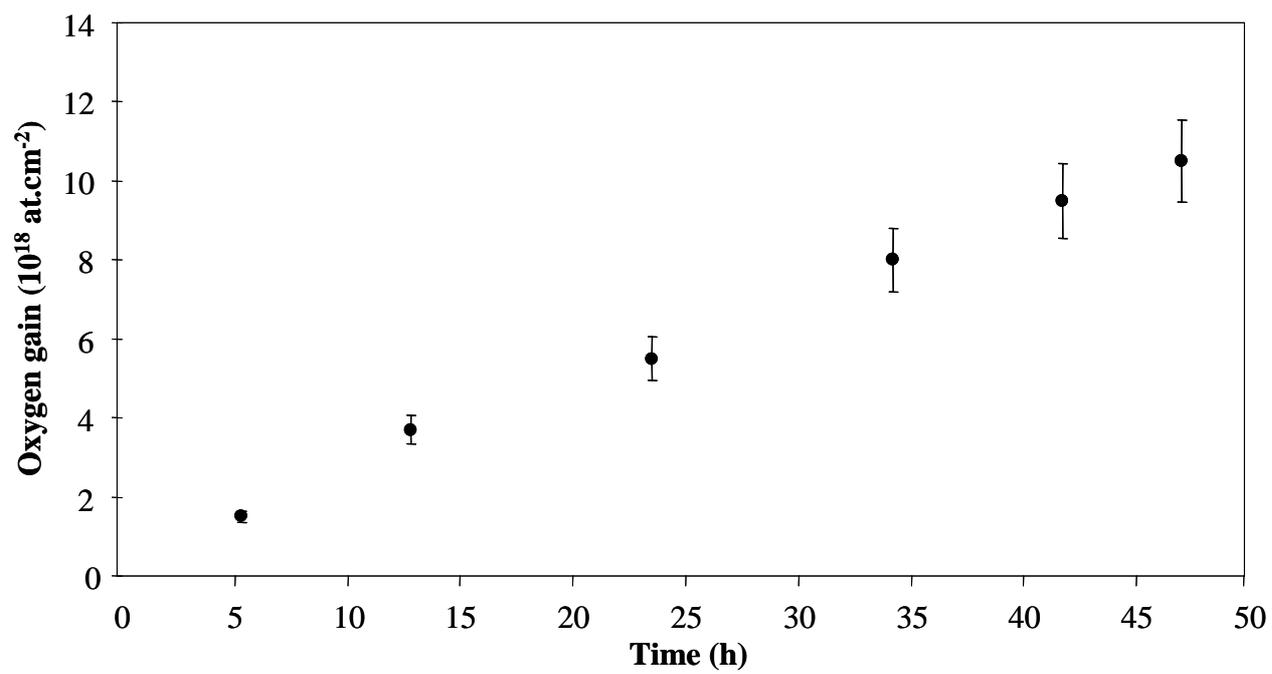

Figure 6

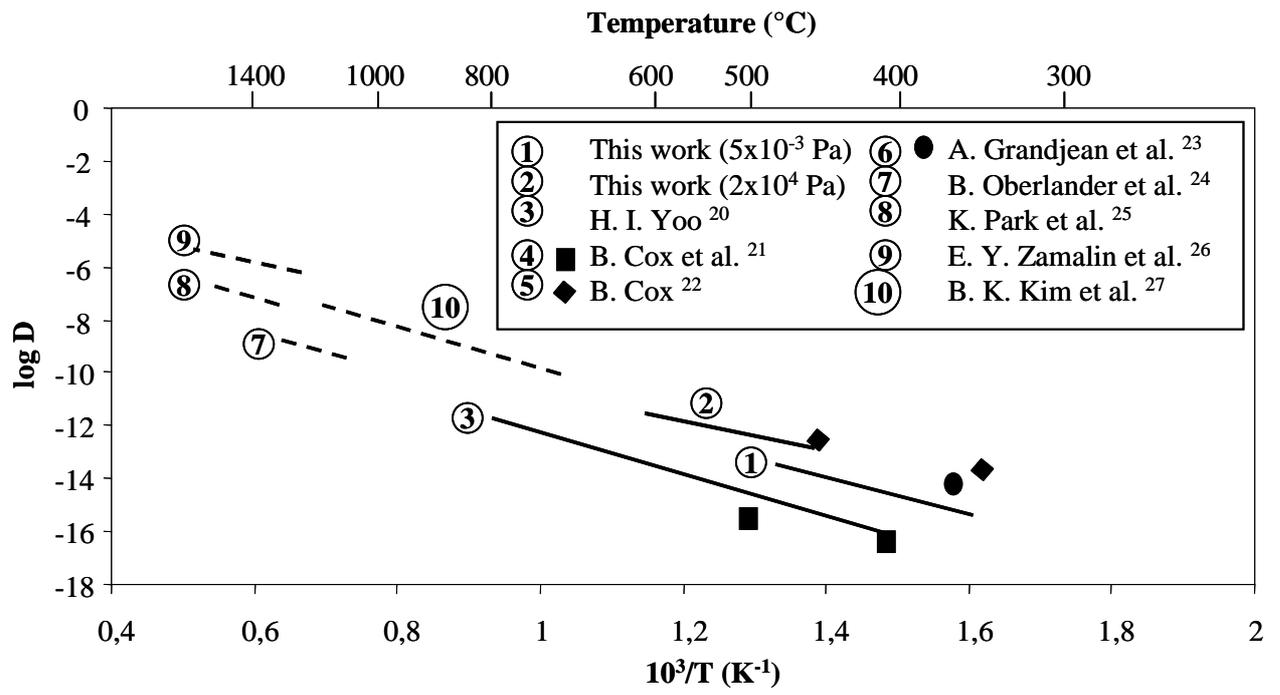

Figure 7

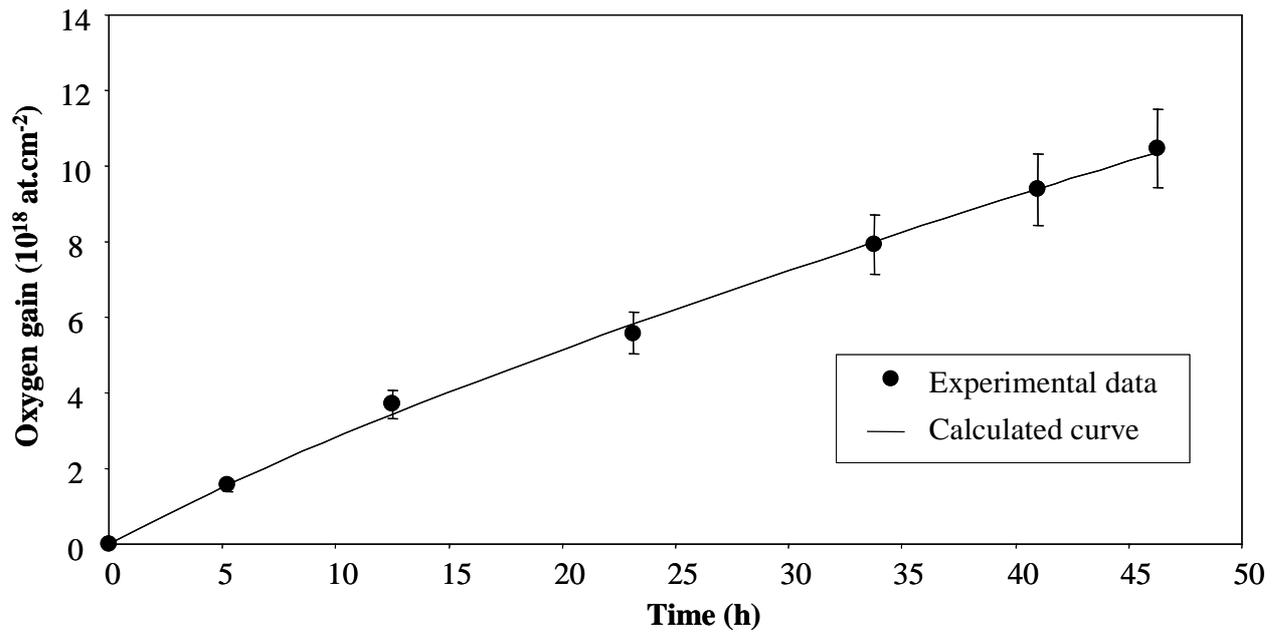

Figure 8

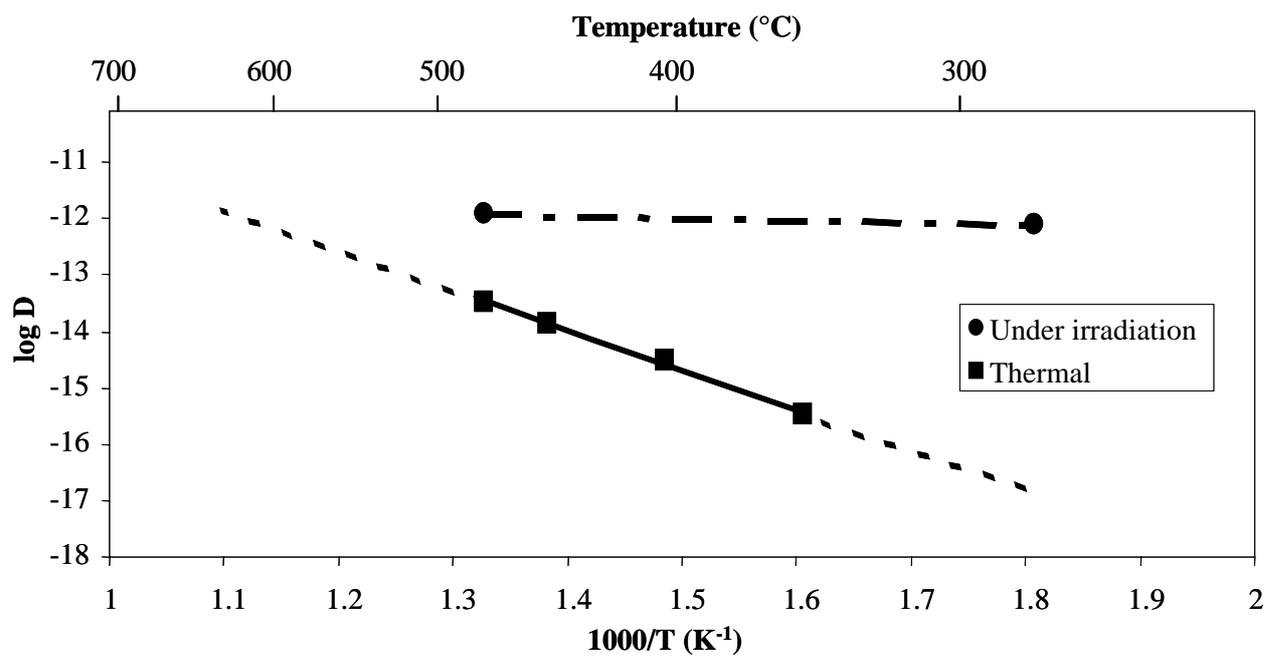

Figure 9

| Irradiation time (hours) | Fluence ($10^{15}$ at.cm$^{-2}$) | Oxygen gain G ($10^{18}$ at d'O.cm$^{-2}$) | ZrO$_2$ oxide thickness (nm) | % tetragonal phase |
|---|---|---|---|---|
| 5.3 | 0.5 | 1.5 | 270 | 75 |
| 12.8 | 1.2 | 3.7 | 680 | 50 |
| 23.5 | 2.2 | 5.5 | 1000 | 52 |
| 34.2 | 3.2 | 8.0 | 1460 | 51 |
| 41.7 | 3.9 | 9.5 | 1740 | 53 |
| 47.0 | 4.4 | 10.5 | 1920 | 56 |

Table 1

| Irradiation time (hours) | Fluence ($10^{15}$ at.cm$^{-2}$) | Oxygen gain G ($10^{18}$ at d'O.cm$^{-2}$) | ZrO$_2$ oxide thickness (nm) | % tetragonal phase |
|---|---|---|---|---|
| 2.7 | 0.25 | 1.0 | 180 | 100 |
| 10.1 | 1.0 | 2.4 | 440 | 100 |
| 11.8 | 1.1 | 2.7 | 490 | 100 |
| 12.8 | 1.2 | 3.1 | 565 | 100 |
| 16.0 | 1.5 | 3.4 | 620 | 100 |
| 33.1 | 3.4 | 6.8 | 1240 | 100 |

Table 2

| $P_{O_2}$ (Pa) | Temperature (°C) | D (cm$^2$.s$^{-1}$) | D$_0$ (cm$^2$.s$^{-1}$) | Ea (eV.at$^{-1}$) |
|---|---|---|---|---|
| 2x10$^4$ | 450 | (1.3±0.3)x10$^{-13}$ | (6±1)x10$^{-6}$ | 1.1±0.2 |
| | 500 | (3.0±0.6)x10$^{-13}$ | | |
| | 600 | (2.7±0.5)x10$^{-12}$ | | |
| 10$^{-3}$ | 350 | (3.4±0.7)x10$^{-16}$ | (8±1)x10$^{-5}$ | 1.4±0.2 |
| | 400 | (3.0±0.6)x10$^{-15}$ | | |
| | 450 | (1.2±0.2)x10$^{-14}$ | | |
| | 480 | (3.3±0.7)x10$^{-14}$ | | |

Table 3

| Temperature (°C) | K (cm.s$^{-1}$) | D* (cm$^2$.s$^{-1}$) | D$_{th}$ |
|---|---|---|---|
| 280 | 1.5x10$^{-9}$ | 7.7x10$^{-13}$ | 1.3x10$^{-17}$ |
| 480 | 1.7x10$^{-9}$ | 1.2x10$^{-12}$ | 3.3x10$^{-14}$ |

Table 4